# Fully-integrated multipurpose microwave frequency identification system on a single chip

Yuhan Yao[1], Yuhe Zhao[1], Yanxian Wei[1], Feng Zhou[1], Daigao Chen[2], Yuguang Zhang[2], Xi Xiao[2], Ming Li[3], Jianji Dong[1, *], Shaohua Yu[2], and Xinliang Zhang[1]

[1]Wuhan National Laboratory for Optoelectronics, Huazhong University of Science and Technology, 430074, Wuhan, China.

[2]State Key Laboratory of Optical Communication Technologies and Networks, Wuhan Research Institute of Posts Telecommunications, Wuhan, China

[3]State Key Laboratory on Integrated Optoelectronics, Institute of Semiconductors, Chinese Academy of Sciences, Beijing 100083, China

**ABSTRACT: Photonic-assisted microwave frequency identification system has been extensively explored and widely applied for civil and defense applications. Fully integration of this system is a promising candidate in attempts to meet the requirements of reduction in size, weight and power. Although many integrated photonic chips have been reported in different technologies, none has monolithically integrated all the main active and passive optoelectronic components. Besides, previous approaches could either identify a single frequency**

**signal instantaneously or identify multiple frequency signals statistically, but not both, hindering the practical applications. Here, we report the first demonstration of fully-integrated multipurpose microwave frequency identification system on a single chip. All critical components including a modulator, a sweeping-frequency filter, a frequency-selected filter, an amplitude comparison function, and parallel photodetectors, are monolithically integrated on the silicon-on-insulator platform. Thanks to the multipurpose features, the chip is able to identify different types of microwave signals, including single-frequency, multiple-frequency, chirped and frequency-hopping microwave signals, as well as discriminate instantaneous frequency variation among the frequency-modulated signals. The frequency measurement range is from 10 to 20 GHz with a measurement error of 409.4 MHz. This demonstration exhibits fully integrated solution and fully functional microwave frequency identification, which can meet the requirements in reduction of size, weight and power for future advanced microwave photonic processor.**

## Introduction

The capability of identifying the frequency of unknown intercepted microwave signals is highly desirable in modern electronic countermeasure, radar warning and electronic intelligence systems[1,2]. With the increasing demand for capacity and the deepening of intelligent development, microwave identification system requires smaller size, larger bandwidth, higher resolution, and lower latency. For this purpose, numerous photonic-based microwave frequency identification methods have been reported in different technologies[3-6]. In particular, leveraging photonic integration approach allowed a dramatic reduction in size, weight and power (SWaP) of the microwave photonic (MWP) systems with improved robustness and complexity, which is of critical importance for wireless and airborne applications[7-11].

Integrated MWP, converges radio-frequency (RF) engineering and photo-electronics, aiming at developing high-performance chip-scale photonic integrated circuit for generation, processing, measurement, and distribution of microwave signals[12]. Complex functionalities have been theoretically and experimentally investigated by integrated MWP. Milestone demonstrations include heterogeneously integrated high-precision optical-frequency synthesizer on III–V/Silicon platform[13], monolithic integrated MWP filters on indium phosphide platform[14] and programmable photonic integrated signal processors[15-18]. However, in the field of microwave frequency identification, fully integration of the multi-functional system is not in sight yet.

The principle of photonic-assisted microwave frequency identification generally involves monotonic mapping mechanism between two parameters, i.e., mapping the signal frequency to a more easily measurable quantity such as power and time. The frequency-to-power mapping (FTPM) scheme can implement instantaneous frequency measurement (IFM) of unknown signal by constructing an amplitude comparison function (ACF). Since the first photonic IFM system was reported in 2006[19], various researches into FTPM-based photonic frequency measurement systems have been demonstrated to achieve an ACF with wider frequency range and higher resolution[20-23]. Recently, integrated approaches based on silicon-on-insulator (SOI) microring resonator (MRR)[24-26], microdisk[27,28], Fano resonator[29], Bragg gratings[30], and indium phosphide Mach-Zehnder interferometer (MZI)[31] have been reported, showing improved performance in bandwidth, robustness and reduction of SWaP. However, only passive optical circuits are integrated, and all the active optoelectronic circuits, such as electro-to-optic conversion and optic-to-electric conversion are implemented with off-chip devices. In addition, these approaches are unable to distinguish multiple-frequency signals simultaneously. By contrast, the frequency-to-time mapping (FTTM) scheme has the ability to achieve multiple frequency measurement (MFM) in a statistical way but not in an instantaneous way. In FTTM schemes, the scanning frequency measurement can be implemented in a dispersive medium[32], a recirculating frequency shifting loop[33], a frequency scanning Fourier domain mode-locked optoelectronic oscillator (OEO)[34-36] or optical filters based on fiber Bragg grating[37], MRR[38,39], stimulated Brillion scattering (SBS) effect[40-42]. However, most of these schemes suffered from either low resolution, bulky system (i.e., only partial integration) or serious sensitivity to environment. Besides, they would lose certain instantaneous frequency of a frequency-agile microwave signal due to the inherent scanning characteristics. In practical electromagnetic environments, the microwave signals are far more complex such as multi-tone, chirped[43], frequency-hopping[44] microwave signals and even their combinations, which is ubiquitous in radar systems and modern communications. Therefore, until now, it is still a great challenge to instantaneously measure the unknown multiple microwave types in complex electromagnetic environments. It is desirable to develop a multipurpose frequency identification system that can identify different types of microwave signals simultaneously and identify the frequency varying instantaneously.

In this work, we report the first demonstration of a multipurpose frequency identification system on a fully-integrated chip, which can identify different types of microwave signals, including single-frequency, multiple-frequency, chirped and frequency-hopping microwave signals, as well as discriminate instantaneous frequency variation among the frequency-modulated signals. By combining their complementary features of the IFM and MFM techniques, a scanning MRR and an asymmetric MZI are employed to realize FTTM and FTPM. All critical components are monolithically integrated on the SOI platform including a dual-parallel Mach-Zehnder modulators (DPMZM), a sweeping-frequency filter, a frequency-selected filter, an amplitude comparison function, and parallel photodetectors (PDs). The frequency measurement range is from 10 to 20 GHz with a measurement error of 409.4 MHz. This demonstration exhibits fully integrated solution and fully functional microwave frequency identification, marking a solid step towards an on-chip universal frequency measurement system and meet the requirements in the pursuit of future advanced MWP processor with a small size, light weight and low power consumption.

## Results

### Chip structure and operating principle

The proposed multipurpose microwave frequency identification chip consists of four main building blocks as illustrated in Fig. 1a. Light generated by a tunable laser source (TLS) is injected to an on-chip DPMZM. By introducing unknown RF signals with a

90° relative phase shift on each MZM electrode and then setting the modulator at the quadrature bias point (90°), an optical single-sideband (OSSB) modulation with carrier-suppression can be generated[45,46]. This signal is split by a 1×2 multimode interferometer (MMI) and then sent to a thermally tunable MRR in one path for microwave frequency classification through FTTM technique. The silicon MRR has a large free spectrum range (FSR) of 80 GHz with a quality factor of ~$2.2×10^5$. Once driving the MRR by a periodic sawtooth voltage, its resonant wavelength will correspondingly experience a periodic redshift, exhibiting a periodic scanning filter. Since the resonance shift is proportional to the power of driving signals, the resonance shift will be quadratic to the voltage over time. When the scanning filter is aligned with the sidebands, a temporal pulse will be observed at the oscilloscope. Therefore, different frequency components can be mapped to different pulses at specific time. In the experiment, the device is temperature-stabilized at 20° by a thermoelectric cooler (TEC) temperature controller.

In another path, three cascaded MRRs constitute a reconfigurable microwave photonic filter, where the unwanted signals can be separated from microwave signals. Its bandwidth and central wavelength are tunable by simply adjusting the direct current (DC) control voltages of each MRR respectively. Subsequently, an IFM block is employed to identify the dynamical frequency variation among the microwave signals. The principle of the IFM is based on an asymmetric MZI, followed by a 2×2 MMI and two on-chip PDs, where the two separated frequency responses are used to define the ACF. The MZI has an FSR of 144 GHz and the extinction ratio is larger than 18 dB. The marks "A"- "D" in Fig. 1a represent the output spectra of input laser, DPMZM, scanning filter, band-selected filter respectively. The unknown microwave signals possibly contain multiple frequencies (represented by $f_1$, $f_2$ and $\Delta f$) and $f_0$ is the optical carrier frequency.

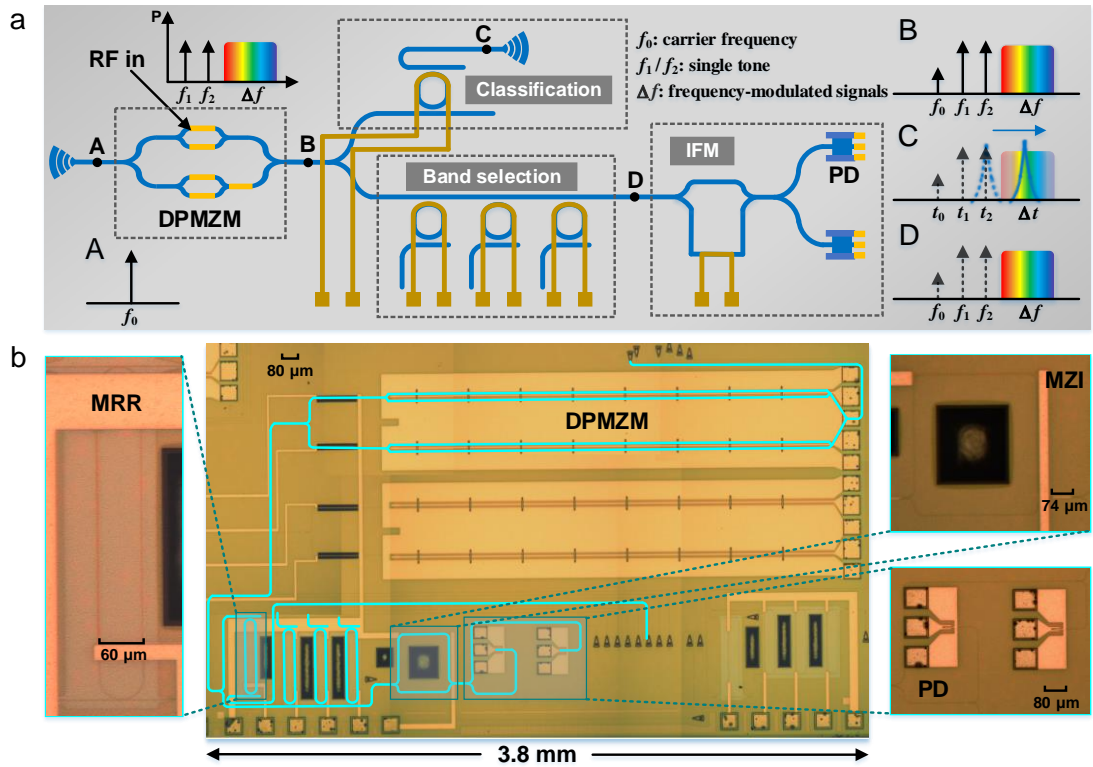

**Figure 1** Monolithic multipurpose microwave frequency identification chip. (a) Schematic diagram of the microwave frequency identification chip. The marks "A"- "D" in Fig. 1(a) represent the output spectra of input laser, DPMZM, scanning filter, band-selected filter respectively. (b) Microscope image of the whole fabricated chip as well as close-ups of fundamental photonic components such as the MRR, MZI and PD. (DPMZM: dual-parallel Mach-Zehnder modulator; RF: radio frequency; PD: photodetector; $f_0$: carrier frequency; $f_1/f_2$: single tone; $\Delta f$: frequency-modulated signals.)

We emphasize that this design allows us to overcome the limitations of existing IFM or MFM systems. The chip can not only identify different kinds of microwave signals, but also monitor the frequency-varying signals in a dynamic manner. The chip was fabricated on an SOI wafer with a silicon layer thickness of 220 nm and a buried oxide (BOX) layer thickness of 3 μm using CMOS-compatible processes. Detailed descriptions can be found **in Methods**. The fabricated chip was wire-bonded to a print circuit board (PCB). Figure 1b shows the microscope image of the fabricated chip. The chip contains four identical MRRs and an MZI, as well as active optoelectronic elements. Some zoom-in photos of the MRR, MZI and PD are also shown. The MRR consists of a race-track ring resonator, two straight waveguides. One of the race-track is integrated with a TiN microheater. Deep air trenches are positioned between the adjacent MRRs and MZI to suppress the thermal crosstalk. The radius of the half-ring is 30 μm to reduce the footprint. The widths of the half-rings and the straight waveguides are 500 nm to guarantee fundamental mode transmission. The width of the race-track region is set as 2 μm to reduce the scattering loss[47]. A linear adiabatic taper with the length of 40 μm connects the half-ring and the race-track to reduce the coupling loss and convert the TE mode from the fundamental mode waveguide to the multi-mode waveguide. The chip footprint is 3.8 mm×2.2 mm. The total chip weighs 6 g after packaging.

**Calibration of the multipurpose system**

To further illustrate how the system collaborates in real measurement, we first calibrate the key elements (i.e., the MRR and MZI) of two schemes. Figure 2a shows the normalized transmission spectra of the MRR under different DC voltages. The corresponding resonant wavelength shift has a quadratic function relationship with the loaded voltage as expected, shown by the red fitting curve in Fig. 2b. We further use a frequency swept laser source and a high-resolution power meter to characterize the Q factor of the MRR. The measured 3 dB-bandwidth is 7 pm at wavelength of 1547.482 nm, resulting a Q factor of ~$2.2\times10^5$. In the experiment, a periodic sawtooth voltage is applied on the MRR, ranging from 0 V to 4 V with a period of 250 ms, guaranteeing a tunable frequency span of over 30 GHz. The response time of thermal heater on the MRR is approximately 82 μs. We scan a given microwave frequency from 10 to 20 GHz to determine the relationship between frequency and time. The output signal was amplified with an erbium-doped fiber amplifier (EDFA), then detected by a photodetector (Discovery, DSC-40S). A 200 MHz oscilloscope (OSC, RIGOL, DS4022) was used to record the appearance time of temporal pulse. Thus a quadratic function is determined between the frequency and the pulse delay, which is the lookup table to estimate the unknown frequency as shown in Fig. 2c. The FTTM-based scheme can identify different types of microwave signals. For many applications, the temporal frequency profile of the frequency-varying signals necessitate to be tracked or identified in a real-time manner. Considering a case that we need to know the dynamic frequency changes of the time-varying signals and block an unwanted jamming signal, this is where the IFM block comes into force. We employ FTPM to realize IFM after microwave photonic filtering. Three cascaded MRRs constitute a bandstop filter, where the central wavelength and bandwidth are adjustable by simply changing the applied voltage of each MRR. An asymmetric MZI with complementary transmission responses is employed to establish ACF. Figure 2d-e show the transmission spectra of two output ports of the MZI. The ACF in Fig. 2f is the power ratio of the two output ports. Therefore, a fixed relationship between the input frequency and the optical powers is established.

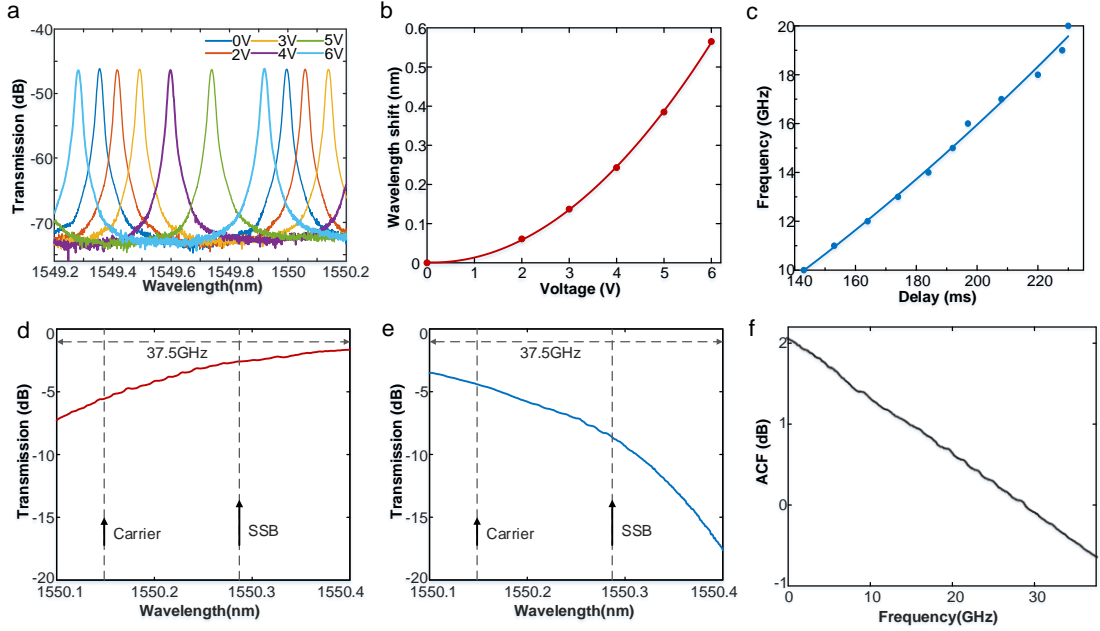

**Figure 2** Characteristic of the scanning filter and the asymmetry MZI. (a) The spectral response of the MRR at different DC voltage. (b) The wavelength shifts as a function of the loaded voltage. (c) Function between the microwave frequency and the time delay. (d, e) Optical transmission spectrum of the two ports of the MZI. (f) ACF curve.

We first verify the system performance by measuring single frequency signals. Figure 3a shows the microwave frequency estimated using the MRR versus the actual input frequency for each of the test tones, as well as the measurement error. It illustrates that the system has the ability of broadband microwave frequency measurement from 10 GHz to 20 GHz with a root-mean-squares error of ~409.4 MHz. With an OSSB modulation, the ideal measurement range will be only determined by the FSR of the MRR, which is 80 GHz in this case. In practice, the on-chip DPMZM has a 3 dB-bandwidth of 22 GHz, representing the upper limitation of the effective measurement range. Besides, carrier-suppressed OSSB modulation is not perfect due to the power and phase mismatches of RF amplifiers. At the same time, the large bandwidth of scanning filter determines the lower limitation of the microwave frequency range. The measurement error is mainly due to the insufficient resolution of the MRR, which can be further improved by using an ultrahigh-Q filter[48]. Figure 3b shows the estimated microwave frequency by inverting the ACF curve with a root mean square error of ~483.8 MHz. The frequency uncertainty is due to the uneven distribution of the on-chip DPMZM and the amplified spontaneous emission noise induced by EDFA to compensate for the link loss.

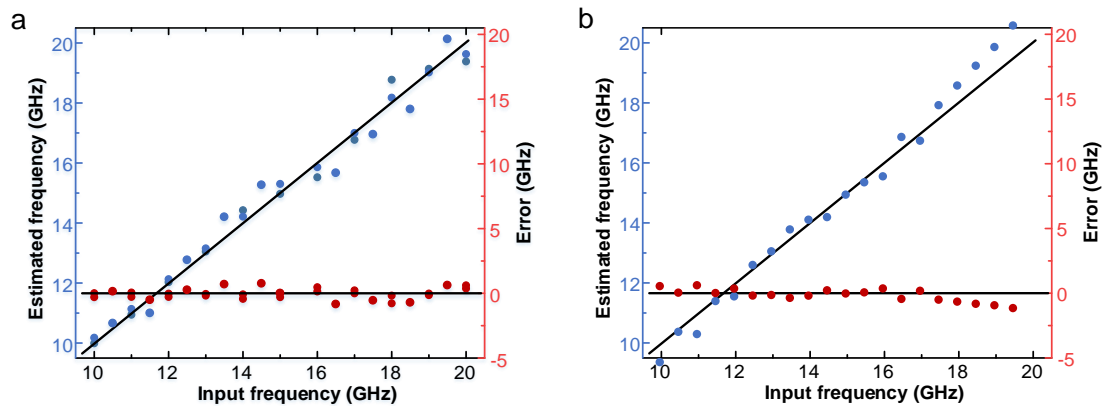

**Figure 3** Estimated frequency (blue dots) and corresponding error (red dots) measured by the FTTM (a) or the FTPM scheme (b).

**Classification of different types of signals**

The FTTM-based scheme is able to identify multi-tone microwave signals and discriminate different types of signals in an easily observable way. To evaluate the performance of the multiple frequency measurement, we prepare several samples to be measured with an arbitrary waveform generator (AWG). Different random combinations of two-tone signals are loaded onto the DPMZM, set as 10 and 15 GHz, 11 and 16 GHz, 12 and 17 GHz, 10 and 11 GHz, respectively. Unambiguous emerging pulses can be observed on the oscilloscope as shown in Fig. 4. The actual measured values are labeled in the diagram and the measured deviation is less than 510 MHz for each tone. It is noted that the amplitude of the temporal pulses gradually degrades as the frequency increases. This is owing to the loss of the on-chip modulator at higher frequency band. Since it is a fully integrated chip, on-chip optical amplification is not available to compensate for the link loss. The two tones in Fig. 4d are just distinguishable, thus the frequency resolution is ~1 GHz. It is slightly larger than the 3 dB bandwidth of the MRR, which is around 875 MHz.

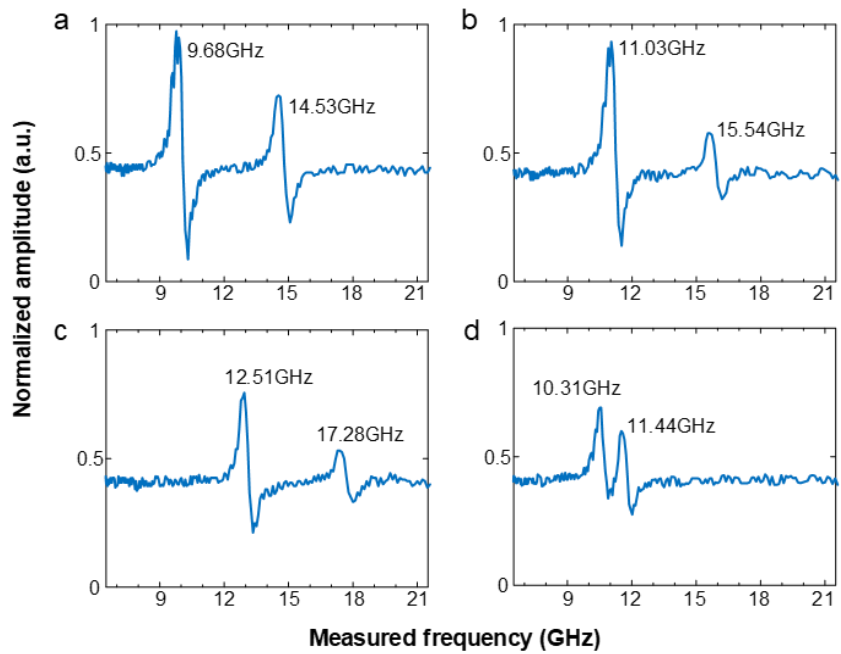

**Figure 4** Measurement results of multiple frequency signals with different input frequencies. (a) 10, 15 GHz. (b) 11, 16 GHz. (c) 12, 17 GHz. (d) 10, 11 GHz.

Radio-frequency chirped waveforms are extensively used in many applications such as radar systems. The ability to acquire the frequency information of chirped microwave signals is an indispensable requirement before performing signal analysis and countermeasure. Based on FTTM, our proposed system is capable of identify the chirped signals. In the experiment, chirped microwave pulses are generated by the RF AWG. The pulse width and repeat interval are set as 1.6 μs and 4 μs, respectively. At a random time during the scanning period, the recorded power is related to the deviation between the instantaneous input frequency and the instantaneous scanning frequency. Thus, the measured waveform is an intensity envelope with random power values filled. Figures 5a, c show the measured results of our method, in which chirped waveforms centered at 15 GHz with spans of 4 and 6 GHz is recorded. Figures 5b, d are the measured results by a commercial electrical spectrum analyzer (ESA) for comparison. It can be seen that the measured frequency span matches well with the input frequency span but we still cannot tell the frequency at specific time. Here we define the frequency span as the span between the pulses when the power is 10% above the noise floor. The root-mean-squares error of the measured bandwidth is 2.12%.

Estimating and tracking the parameters of frequency hopping spectrum-spread signals are important tasks with applications in both civilian and military domains[49]. The problem is particularly challenging when the hopping frequencies are unknown. In addition to dwell frequency, hop timing is randomized as well for added protection. In view of this situation, here we first identify the frequency information of the frequency hopping signals based on FTTM. Similar to the chirped microwave signal, the measured frequency hopping signal exhibits an envelope filled with random power values as it is also a time-varying signal. **See Discussion** for detailed characteristics of different waveforms. Programmed by the AWG, a group of frequency hopping signals with different frequency steps are loaded onto the on-chip DPMZM and the duration for per tone is 80 ns. Figures 5e-h show the measurement results of our method versus the ESA measurement. Here we take the peak value of the envelope as the measuring point. It can be seen that the measured results are aligned well with the actual status and the measurement error is about 166.9 MHz. In Fig. 5g, a frequency hopping sequence containing frequency components of 10, 13, 15 and 17 GHz is observed successfully but we cannot know the chronological order of those frequencies. For many applications, the temporal frequency profile of the frequency-varying signals necessitate to be tracked or identified in a real-time manner. In the next section, we will show how to identify the different frequency-varying signals dynamically and instantaneously.

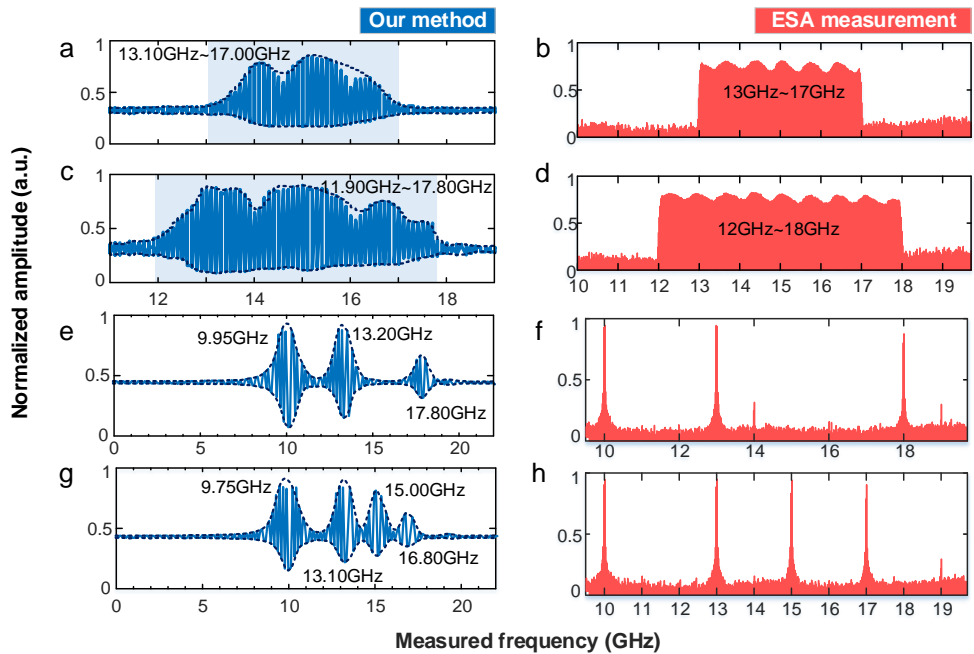

**Figure 5** Measurement results of chirped frequency signals with different chirped band and frequency-hopping signals. (a, c, e, g) are system measurement results. (b, d, f, h) are ESA measurement results. The chirped pulses are centered at 15 GHz with different bandwidth of 4 and 6 GHz. The set of hopping frequencies is 10, 13, 18 GHz and 10, 13, 15, 17 GHz.

**Filtering and dynamic frequency identification**

To show that the system can identify the temporal frequency profile of unknown, rapidly varying RF signals in a dynamic manner, a series of chirped frequency signals and frequency-hopping signals are generated and fed to the system respectively. Figures 6a, d show the spectrogram of the unknown injected RF signals. The on-chip PD has a

3 dB-bandwidth of 33 GHz. It is important to note that since only one RF probe was available for the integrated PDs in the experiment, only one output port was employed in actual tests. As a result, the corresponding transmission response, similar to the blue curve in Fig. 2e, was used instead of the ACF to discriminate different frequencies. Finally, the output power is recorded by an OSC, the frequency information is mapped into the amplitude as shown in Fig. 6b, e.

For the chirped frequency signals, the pulse width and repeat interval are set as 160 ns and 200 ns. The resulting instantaneous frequency profile is extracted as shown in Fig. 6c, covering a span of 12-18 GHz. Here we set the extracted amplitude corresponding to 20 GHz as an upper limit. The signal higher than 20 GHz is not included in the lookup table hence classified as noise. The shadow represents the noise floor, a minimum detectable threshold of the extracted instantaneous amplitude, which is determined by the system bandwidth. For the frequency-hopping signals, the amplitudes of the different frequency bursts are equal and the hopping interval is 80 ns. Suppose that there is an undesired frequency of 10 GHz in the RF sequence, which can be eliminated by properly adjusting the frequency-selected filter. After filtering out the 10 GHz signal in the experiment, its amplitude is much lower than the output amplitude of the 20 GHz frequency and is therefore classified as noise. It can be seen that there is still a weak signal in the shaded area which happens to be the previously filtered out jamming signal. Finally, by using the inverse ACF mapping, the instantaneous frequency profile could be successfully extracted, as shown in Fig. 6f. Frequency components at 13, 15 and 17 GHz has been reconstructed instantaneously while 10 GHz is filtered out.

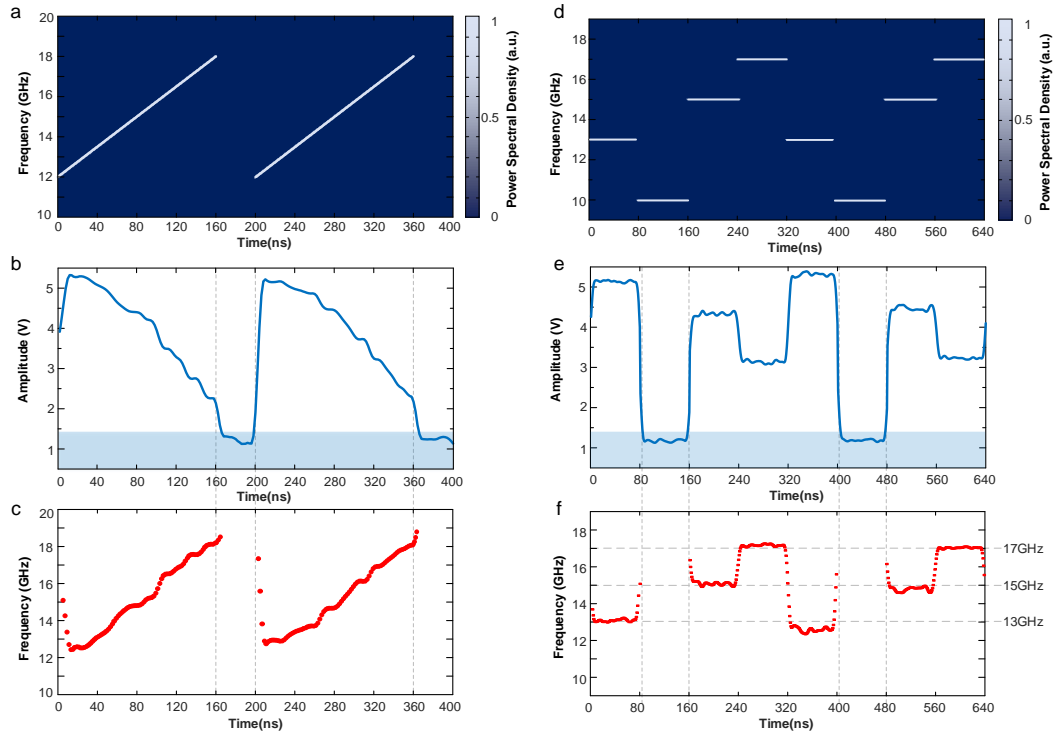

**Figure 6** Experiment of dynamic frequency identification. (a, d) Spectrogram of the input unknown chirped frequency and frequency-hopping signals. (b, e) The instantaneous waveform of chirped frequency signals and frequency-hopping signals recorded by the OSC after filtering. (c, f) The corresponding frequency component in a dynamic manner.

**Discussion**

In the above demonstrations, a powerful and multipurpose microwave frequency identification system is proposed to identify all the main types of microwave signals, i.e., single frequency, multiple-frequency, chirped and frequency-hopping microwave signals. The measured waveforms of these four types have distinguishable characteristics, which are summarized in Table 1. Different waveforms can be easily classified through FTTM technique. Time-invariant signals and frequency-modulated signals can be distinguished explicitly by judging the measured pulse being filled with random power or not. For the time-invariant signals, we can check whether the signal is single frequency or multiple-frequency by counting the number of pulses. For the frequency-modulated signals, we can further distinguish them by judging whether the pulse envelope is continuous or discrete. In addition, the temporal frequency profile can be extracted through FTPM technique where the frequency variation is derived in a dynamic manner. Thus our system has the ability to identify diverse microwave signals and discriminate the frequency variation instantaneously.

Table 1 | Classification criterion of measured microwave signals

| Microwave type | Pulse filled | Envelope | Typical waveform (statistical) | Typical waveform (instantaneous) |
|---|---|---|---|---|
| Single frequency | No | Single | 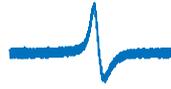 | 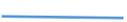 |
| Multiple-frequency | No | Multiple | 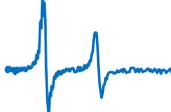 | Not available |
| Chirped frequency | Yes | Continuous | 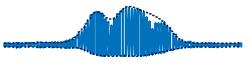 | 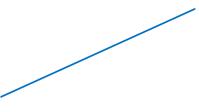 |
| Frequency-hopping | Yes | Discrete | 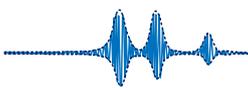 | 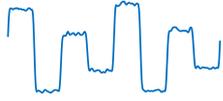 |

The fully-integrated chip is compelling and urgent for a wide range of applications especially for platforms requiring the reduction of SWaP, such as satellites and unmanned aerial vehicles. The fabricated chip has a small size of 3.8 mm×2.2 mm, a light weight of only 6 g. The total power consumption of the system is estimated to be 3.52 W, including the power consumption of the microwave amplifier, sweeping-frequency filter, frequency-selected filter, and PD. Since the chip is fully integrated, link loss is rather critical for overall performance. Active devices like modulators will bring about an undesirable loss of optical power, resulting in a relatively poor signal quality, especially at higher frequencies. Such a high loss in high-frequency band ultimately limits the measurement range. In the future, the chip can be further improved by optimizing the loss of the modulator and integrating on-chip optical amplifiers to compensate the insertion loss of the chip[50].

In conclusion, we have reported the design, fabrication and experimental characterization of a fully-integrated multipurposemod microwave frequency identification system on the SOI platform, leveraging the maturity of the CMOS fabrication process. The chip includes a modulator, a sweeping-frequency filter, a frequency-selected filter, an MZI and parallel PDs. To the best of our knowledge, this is the first monolithically integrated frequency identification system reported to date. Four types of typical microwave signals in terms of single frequency, multiple frequency, chirped and frequency hopping microwave signals can be classified according to the feature of measured waveforms. Chirped frequency signals and frequency-hopping signals are also identified in a dynamic manner. The measurement range is from 10 to 20 GHz with a resolution of 1 GHz and a measurement error of 409.4 MHz. Such a chip has the advantages of small size (3.8 mm×2.2 mm), light weight (6 g) and low power consumption (3.52 W), which is promising for applications in radar and electronic warfare systems. The successful implementation of our microwave frequency identification system on a chip marks a significant step forward in the full integration of active and passive components on a single chip and opens avenues toward real applications of miniature on-chip MWP systems.

**Methods**

**Device fabrication.** The chip was fabricated on a commercial standard SOI wafer with a silicon layer thickness of 220 nm and a BOX layer thickness of 3 μm using CMOS-compatible process. The passive waveguides and grating couplers were first defined and fabricated by deep ultra-violet (DUV) photolithography and the inductively-coupled-plasma (ICP) etching process. The modulator was based on a carrier-depletion MZM. Multiple-step P-type ion-implantations were then used to achieve a uniform background doping concentration of ~$2 \times 10^{17}$ cm$^{-3}$. N-type implantations with a target concentration of $4 \times 10^{17}$ cm$^{-3}$ were then employed to compensate the background doping and form an abrupt PN junction. Highly doped P + + and N + + regions were both located 1 μm away from the waveguide sidewalls and then rapidly annealed for the doping activation. To fabricate the germanium-on-silicon photodetector, a 500 nm-height germanium layer is deposited and pattered on top of the silicon waveguide with its top ion implanted by N+ for ohmic contact. The bottom Si waveguide is implanted by P and P++ for ohmic contact. The cuprum and aluminum electrodes were then deposited to form the electric contact. Finally, isolation trench was defined by a deep dry etching process to provide thermal isolation to adjacent devices.

## Data Availability

The data that support the findings of this study are available from the corresponding author upon request.

## Corresponding Author

*Email: jjdong@hust.edu.cn

## Authors' Contributions

Y.H.Y and Y.H.Z. conceived the ideas and conducted the design. Y.H.Y and Y.X.W performed the experiment measurement. D.G.C and Y.G.Z designed the modulator and photodetector. Y.Y wrote the manuscript. Y.Y, F.Z, J.J.D discussed the results. J.J.D and X.L.Z supervised the study. All authors discussed the data and contributed to the manuscript.

**Funding Sources**

This work was partially supported by the National Key Research and Development Project of China (2018YFB2201901), the National Natural Science Foundation of China (61805090, 62075075).

**Competing Financial Interests**

The authors declare no competing financial interest.